# Subwavelength plasmonic antennas based on asymmetric split-ring-resonators for high near-field enhancements

Yue You, Xiao-Jing Du, Lin Ma, Hua Qiu, Jun He, and Zhong-Jian Yang*

*Hunan Key Laboratory of Nanophotonics and Devices, School of Physics, Central South University, Changsha 410083, China*

*zjyang@csu.edu.cn

**ABSTRACT:** As for plasmonic antenna structures that generate localized near-field enhancement, the most effective current implementations are based on electric dipole resonance modes, but this approach also imposes limitations on their further optimization. Here we introduce an ASRR structure whose ASR mode enables differential charge distribution across both sides of the split. Through asymmetric regulation, charges at one end can become highly localized, thereby achieving efficient near-field enhancement. The formation of this structure was initially driven by a hybrid computational framework integrating evolutionary optimization with residual neural networks, and subsequently simplified into an ASRR prototype using the Occam's Razor principle. The ASRR dimer structure can achieve an electric field intensity enhancement over 6.5 times larger than a traditional nanorod dimer, while maintaining a compact size (<1/3 the working wavelength). The ASRR configuration also demonstrates superior Purcell factor and fluorescence enhancement. These results can find applications in surface-enhanced spectroscopy, nonlinear optics, and quantum light-matter interactions.

The manipulation of light at the nanoscale has revolutionized photonics, enabling precise control over light-matter interactions [1–3]. A key platform for such control is plasmonic nanostructures, which exploit surface plasmon resonances to confine light into subwavelength volumes, surpassing the diffraction limit and achieving exceptional near-field enhancement [3–11]. Advances from simple geometries, such as single particles, to complex configurations, like dimers or nanoparticle-on-mirror cavities, have significantly enhanced electric field intensity enhancements (FIEs), driving breakthroughs in surface-enhanced Raman spectroscopy (SERS), fluorescence enhancement, nonlinear optics, and strong coupling with single emitters [1,5,9,12–17], etc. Consequently, optimizing the FIE at nanoscale has always been a core research focus [4,10,12,18].

One effective strategy for enhancing near-field intensity is to reduce the gap size in plasmonic nanostructures [18,19]. However, current gap sizes have pushed down to ~1 nm scale, approaching practical and quantum mechanical limits [20,21]. Further reduction hindered by the need of target objects and application need, e.g., quantum dots [22] and fluorescence enhancement [5,16], respectively. While cascading amplification effects for strong FIE can be achieved by integrating large-scale background structures near plasmonic cavities [23,24], this approach often increases system volume and could be incompatible with applications like single-emitter strong coupling [24]. Therefore, optimizing FIE within individual subwavelength structure remains critical for advancing broad practical applications.

Currently, it is challenging to make further breakthroughs in FIE for individual plasmonic antennas as the optimization of the structural morphology is not straightforward. Physically, those plasmonic antennas are mostly based on electric dipole or coupled electric dipole plasmon modes [10,17], and there is no clear physical picture for breaking through the existing framework of the working modes. Recently, numerical optimization methods have emerged for the inverse design of nanophotonic devices [25–30], especially the evolutionary optimization (EO) based on evolutionary algorithms combining with numerical simulations has provided new insights into the optimization of plasmonic structures [26,30]. However, most of the new structures generated in the EO have poor performance, significantly limiting the optimization efficiency due to the considerable simulation time for each structure. Moreover, the complex morphologies resulting from numerical optimization pose challenges for mode analysis, hindering the development of new physical insights and breakthroughs beyond existing frameworks.

In this work, we introduce a novel plasmonic antenna, asymmetric split-ring resonator (ASRR), where the charges on the two sides of the split maintain two distinct types. The structural asymmetry based on the asymmetric split-ring (ASR) mode can induce highly asymmetric charge localization, resulting in a dramatically enhanced near-field response. The ASSR is proposed based on two steps. A complex morphology implicitly related to the ASRR is first deduced through a computational model, integrating EO with a transfer residual convolutional neural network (ResNet) [31], named EoNet. Neural networks have been successfully applied to photonic nanostructure optimizations [32–42]. Utilizing EoNet, we achieve a ~10-fold increase in optimization efficiency compared to pure EO processes. Then, by applying Occam's Razor principle on the above complex structure, an ASRR prototype is obtained. Specifically, the FIE of an ASRR dimer exceeds that of a traditional nanorod dimer (with the same gap size) by a factor of more than 6.5 times. The ASRR framework also significantly outperforms conventional antennas in both fluorescence enhancement and Purcell factor. Remarkably, the size of an ASSR remains comparable to a traditional nanorod dimer, being less than 1/3 of the working wavelength.

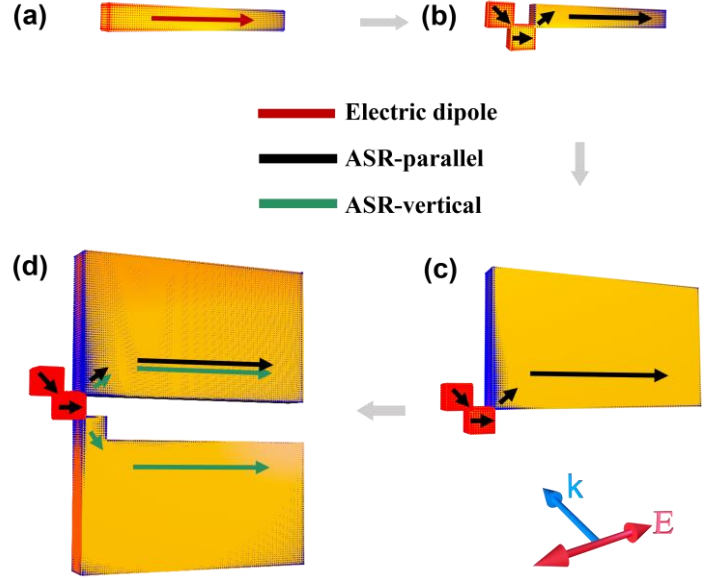

FIG. 1. Schematic diagram of plasmonic structures with ASR modes. Four plasmonic structures and their typical resonance modes under the given excitation are shown. The red and blue colors on each structure are schematic of positive and negative charge distributions, respectively. (a) A single nanorod with a longitudinal electric dipolar mode. (b) An ASRR with a fundamental ASR mode with gap-parallel polarization (ASR-parallel). (c) An ASRR with the tail nanorod being enlarged. (d) A composite ASRR with another additional fundamental ASR mode with gap-vertical polarization (ASR-vertical). The structures in (b), (c) and (d) are named by ASRR-1, ASRR-2 and ASRR-3 for the sake of clarity. The typical current distributions for different modes are schematically illustrated by the arrows with different colors.

Figure 1 schematically illustrates the charge distribution characteristics of the proposed ASRRs. In a nanorod, the structural symmetry leads to a symmetrical charge distribution for the longitudinal dipolar mode (Fig. 1(a)). By introducing a concavity at an off-center position, one breaks this symmetry, transforming it into an ASRR structure (ASRR-1). Figure 1(b) illustrates the typical current and charge distribution of its fundamental mode under gap-parallel polarization (ASR-parallel). The main features resemble those of the well-known symmetric split-ring resonator (SRR) [43–45], with the fundamental mode retaining the characteristic of two opposite charges across the gap.

This charge asymmetry becomes more prominent when the right part of the structure is enlarged (ASRR-2), resulting in enhanced charge accumulation at the tip-cube, as shown in Fig. 1(c). By introducing an additional nanorod beneath the concavity of ASRR-2, a new ASRR structure is obtained (ASRR-3, Fig. 1(d)). ASSR-3 supports another type of ASR mode corresponding to a fundamental split-ring mode excited under gap-vertical polarization [43]. Due to the overall structural asymmetry, this resonance can be classified as an ASR-vertical mode. The cooperative effects between the two ASR modes further enhances charge localization at the tip-cube.

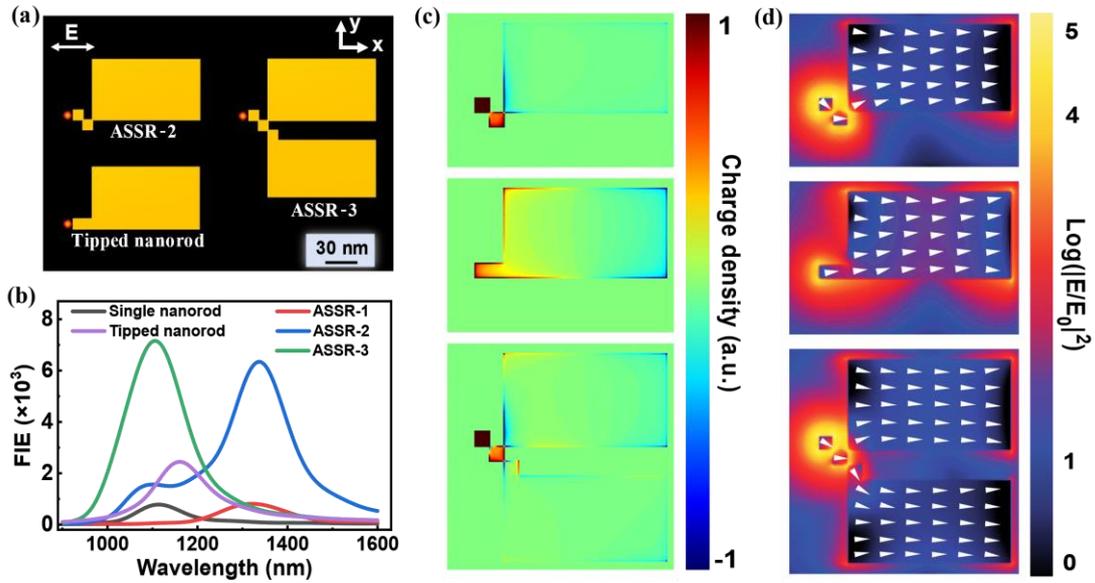

FIG. 2. The FIE properties of ASRRs. (a) The ASRR-2 and ASRR-3 structures. A structure without a concavity is also shown (denoted by "tipped nanorod"). The size of each smallest cube is 10 nm. The height for all structures is 10 nm. The red dots denote the point where FIEs are calculated. (b) FIE spectra for the ASRR-2, ASRR-3 and tipped nanorod. The results for a single nanorod and ASSR-1 are also shown for comparison. The length, width and height of the nanorod are 109, 10 and 10 nm, respectively. (c) Charge distributions for resonant modes of ASRR-2, ASRR-3 and tipped nanorod. (d) Logarithmic FIE and current distributions (white arrows) for resonant modes of ASRR-2, ASRR-3 and tipped nanorod.

The advantages of the ASRR structural configurations are verified by the direct finite-difference time-domain (FDTD) simulations (Fig. 2(a)). To quantify the electric near-field response, we calculate the FIE at a left point 5 nm from the tip-cube for each structure (Fig. 2(b), see also the extinction spectra in Fig. S1 in Supplemental Material). The ASRR-1 structure indeed exhibits two types of charges on both sides of the gap. For ASRR-2 structure, we observe a corresponding increase in FIE around $\lambda$ = 1350 nm. In striking contrast, the removal of the concavity will revert the charge distribution to a conventional dipolar mode, with positive and negative charges symmetrically distributed around the geometric center. Although the presence of sharp tip can enhance charge density, the resulting FIE is significantly weaker than that of ASRR-2 (Figs. 2(b) - 2(d)). A small shoulder in the ASRR-2 spectrum appears around $\lambda$ = 1100 nm, which corresponds to a higher order mode (see Fig. S2).

The ASRR-3 structure consists of two parallel nanorods connected by a linking cube, with an additional cube inserted to suppress direct interactions between the rods. The cooperative effects between the two types of ASR modes lead to a more pronounced asymmetric charge distribution on the tip-cube (Fig. 2(c)). The blue shift of the FIE spectrum arises because the large rectangle nanorods in ASRR-3 also support electric dipole modes (Fig. 2(d)). The interaction between the electric dipole modes on the two rods induces an antibonding hybridization effect, leading to the observed blue shift [12,46]. Considering the wavelength effect, i.e., the blue shift relatively results in smaller FIE (FIE~$\lambda$) [24,47,48], the FIE value of ASSR-3 distinctly outperforms the ASSR-2. Compared to the single nanorod, the ASRR-3 structure achieves a more than 9-fold increase in FIE.

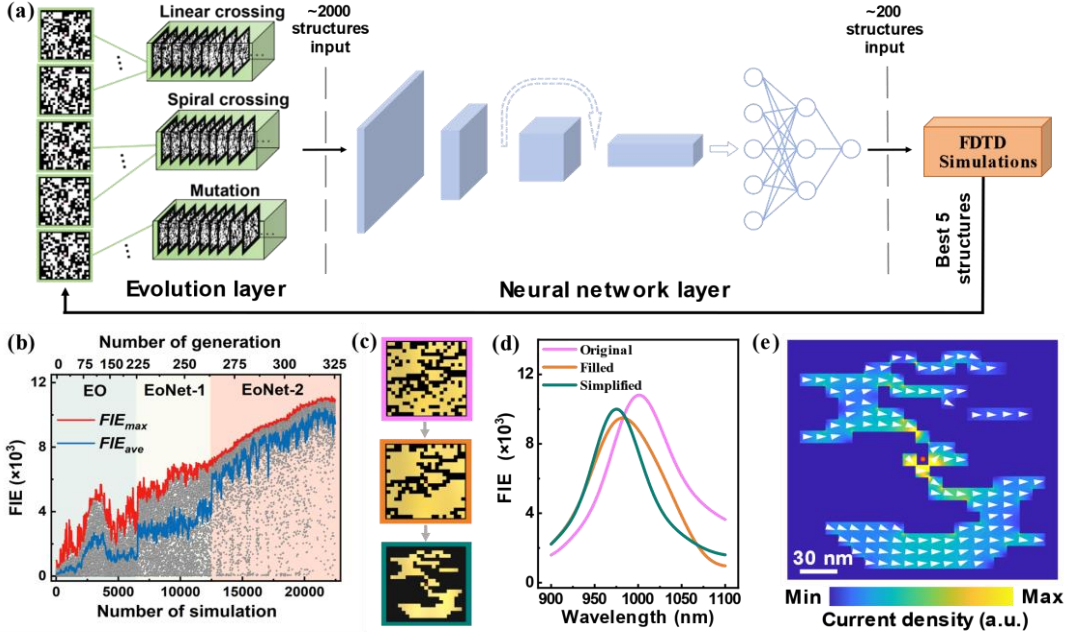

FIG. 3. (a) Schematic diagram illustrating the structural optimization process utilizing the EoNet. The entire process can be divided into three main parts: evolution layer, neural network layer, and FDTD simulations. (b) FIE evolve with number of generation (simulation). The first 225 generations involve pure EO, and the remaining generations involve the EoNet which can be further divided into two stages (EoNet-1 and EoNet-2). Each gray dot represents a simulation, and the maximum and average value of FIE are shown as red and blue lines, respectively. (c) Morphology simplifications by using Occam's Razor principle. The original, filled and simplified structures are shown from top to bottom. (d) FIEs at the reference point for the three structures in (c). (e) Currents distribution of the simplified structure.

We now discuss the structure generation process underlying the ASRR designs. We introduce a computational model, EoNet, which integrates EO as the evolution layer and a transfer ResNet as the neural network layer. By leveraging the neural network layer's recognition capabilities, EoNet efficiently filters out low-performance structures generated by the evolution layer, as illustrated in Fig. 3(a). To train the EoNet program, we first generate ~7,000 structures over 225 generations using EO (more details of creating new individuals in Supplemental Material). Each structure undergoes FDTD simulations, with the calculated FIEs used as training data. The FDTD simulations are conducted based on the FIE at a fixed point, with the surrounding structure discretized into 10 nm cubic units, each either containing gold or vacuum. Here, we select the field enhancement at the geometric center within a spatial volume of 210 × 210 × 10 nm³ as the optimization parameter. Given the vast design space $4 \times 10^{132}$ [26], direct simulations are infeasible. The working wavelength is taken to be $\lambda = 1000$ nm and gold's optical properties are taken from Palik's book [49].

Once the EoNet achieves sufficient recognition accuracy (see Fig. S3 and related discussion in the Supplemental Material), we employ it to screen structures generated by the evolution layer. For each generation, neural network layer can effectively eliminate approximately 90% of the candidate structures. The remaining structures undergo FDTD simulations, significantly reducing the optimization time. Then, the top 5 structures are used to generate 2000 structures for the next generation. Figure 3(b) illustrates the structural evolution across iterations, which can be divided into two distinct phases. In the initial phase (~7,000 samples), optimization relies solely on EO, resulting in a significant gap between the average ($FIE_{avg}$) and maximal ($FIE_{max}$) values. This discrepancy highlights the inefficiency of pure EO, as many low-FIE structures consume computational resources without contributing to performance improvement.

Starting from the ~7,000th structure, the optimization process integrates EoNet, leading to a significant jump in both $FIE_{max}$ and $FIE_{avg}$. A second upgrade to EoNet, implemented via data augmentation at the ~12,500th structure (see Fig. S4 and related data augmentation discussion in the Supplemental Material), further improves recognition performance (EoNet-2). This enhancement extremely narrows the distance between $FIE_{max}$ and $FIE_{avg}$, with both metrics increasing steadily, indicating minimal computational waste. After ~22,000 FDTD calculations (from a pool of over 200,000 candidate structures) across approximately 325 generations, we achieve a FIE of ~11,000. The FIE stabilizes around this value, suggesting diminishing returns, so we stopped further calculations.

The optimized structure exhibits a chessboard type complex morphology, with many cubes contributing minimally to the FIE. To better analyze the underlying physical modes, we simplify the structure while preserving its field enhancement properties. Starting from the original morphology, we first fill hollow inner regions (with cubes in all four neighboring positions) with Au cubes, which has a negligible impact on the FIE at the reference point (Figs. 3(c) and 3(d)). Next, we apply Occam's Razor principle [50] to systematically simplify the structure: we assume that only one Au cube is reduced as a virtual optimization path and record the optimization parameter values. To minimize computational cost, only edge structures (without cubes in four neighboring positions at the same time) are removed. Among all possible optimization paths, we select the one with the smallest impact on the matrix structure as the real path and proceed to the next round of iterative optimization (see Supplemental Material for a further discussion of the Occam's Razor process). The resulting simplified morphology retains a FIE more than 92.5% of the original one (Fig. 3(d)). As shown in Fig. 3(e), this configuration and its mode responses closely resemble the ASRR-3 structure above.

It is important to note that, although our optimization targets the FIE at a fixed point, the resulting structure exhibits a comparable FIE distributed across a broad region near the tip-cube, as evidenced by the FIE distribution (Fig. S6). This behavior contrasts sharply with conventional dimer structures, where FIE is typically confined to the gap region. Instead, the enhanced field in our structure arises from a widely distributed charge accumulation on the surface of the tip-cube, effectively functioning as an open cavity in terms of FIE. As an open cavity, this ASSR configuration achieves a FIE significantly larger than that of common open cavities (e.g., ~9 times greater than that of a single nanorod), demonstrating its superior performance.

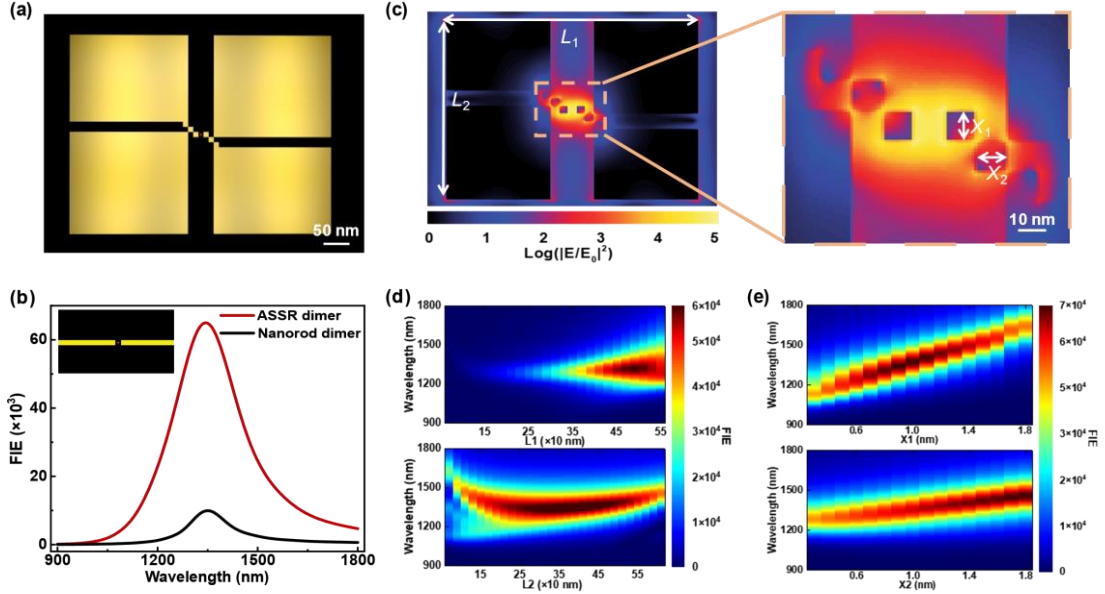

FIG. 4. (a) Schematic diagram of an ASSR dimer. The geometry of the ASSR-3 is the same as that in Fig. 2. The gap size of the dimer is 10 nm. (b) FIE spectrum of the ASSR dimer. The nanorod dimer with the same gap as a reference is also shown for comparison. (c) Logarithmic FIE distribution at the resonance. The right part shows the zoomed view of the gap region. (d) FIE spectra of the ASSR dimer as a function of $L_1$ and $L_2$. (e) FIE spectra of the ASSR dimer as a function of $X_1$ and $X_2$.

Similar to conventional nanorod dimers, we construct a dimer structure based on the ASRR-3 design in Fig. 2. As shown in Fig. 4(a), the ASRR dimer features a 10 nm gap between the tip-cubes, with the FIE at the gap center reaching ~65,000. This dimer is arranged in point-symmetric configuration (the mirror-symmetric configuration shows similar results, see Fig. S7). Remarkably, this structure maintains a compact size, with dimensions less than 1/3 of the working wavelength—comparable to traditional nanorod dimers. To the best of our knowledge, this FIE value surpasses existing subwavelength plasmonic cavities with similar gap characteristics by several times and is more than 6.5 times greater than that of a nanorod dimer with the same gap size (Fig. 4(b)).

Figure 4(c) presents the logarithmic FIE distribution of the ASRR dimer, with the strongest field localization occurring near the tip-cubes (see also linear distribution in Fig. S8). We further investigate the impact of the total lengths $L_1$ and $L_2$ on the FIE behavior (Fig. 4(d)). Within a certain range, increasing $L_1$ is equivalent to increasing the length of the ASRR cavity, leading to an overall redshift of the modes. On the other hand, the antibonding coupling between dipole plasmon modes results in a blueshift. The interplay of these effects stabilizes the resonance wavelength. The variation of $L_2$ shows similar behavior to that above.

The size of the Au cubes near the gap region significantly influences the response wavelength. Increasing either $X_1$ or $X_2$ effectively extends the current path of the ASR mode, inducing a notable redshift (Fig. 4(e)). The FIE exhibits an optimal value as $X_1$ varies: when $X_1$ is too small, the split-ring effect weakens, reducing the ASR mode response; when $X_1$ is too large, the ASR response strengthens, but the charge distribution becomes delocalized across the entire length, leading to poorer locality. The interplay of these competing effects produces the results shown in Fig. 4(e). Similar effects based on the characteristics of ASR can also be found in more different morphologies of the gap region (Fig. S9).

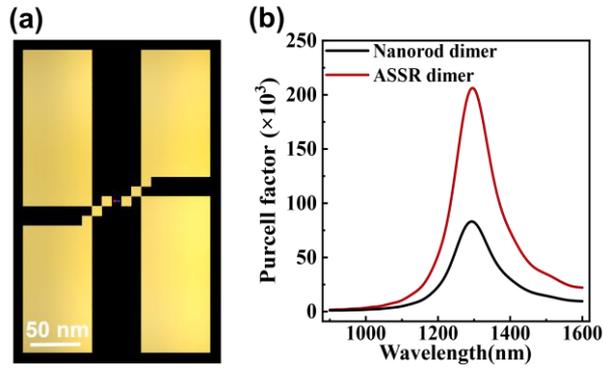

FIG. 5. Purcell factor of an ASSR dimer. (a) Schematic of an ASSR dimer structure, with a dipole source at the gap center. Each small block is a 10 nm cube. (b) Purcell factor spectra of the ASSR dimer and a nanorod dimer with the same resonant wavelength and the gap size.

The ASRR structural configuration also demonstrates significantly superior performance in Purcell factor compared to conventional designs. Fig. 5(a) illustrates an ASSR dimer configuration, featuring a dipole emitter located at the gap center. It achieves a Purcell factor ∼2.5 times higher than that of nanorod dimer, reaching ~200000 (Fig. 5(b)). It is important to note that Purcell factor is not solely determined by FIE [24], which explains why the ASRR's Purcell factor performance and its dependence on structural parameters differ from that of the FIE behavior. This is directly reflected by the fact that the sizes used for FIE and Purcell factor are different (Figs. 4 and 5, see also Fig. S10). Purcell factor serves as a metric for near-field locality, indicating that the ASRR framework offers significant advantages in enhancing cavity-quantum emitter couplings [51,52].

The ASRR dimer also demonstrates significant advantages in fluorescence enhancement compared to conventional nanorod dimers. First, the FIE of the ASSR dimer exceeds that of a nanorod dimer by more than 6.5 times, with a broader spectral response (Fig. 4). Second, the quantum efficiency can reach ~5 times increase over the nanorod dimer (Fig. S11). Combining both factors [5], the ASRR dimer achieves a ~30 times increment in fluorescence enhancement compared to the nanorod dimer.

Conclusion:

In conclusion, an ASRR subwavelength plasmonic structure is proposed based on an EoNet framework. This design transcends traditional electric dipole plasmon resonance by supporting ASR modes with asymmetric charge distribution, which localizes charges at a small tip and significantly enhances the FIE nearby. Our EoNet approach leverages efficient structure recognition to eliminate low-performance candidates, improving optimization efficiency by approximately tenfold compared to pure EO. After screening ~200,000 structures, we identify a high-performance configuration. We further simplified using Occam's Razor principle to derive the proposed ASRR structure with constructive cooperation between two ASR modes. An ASSR dimer structure can achieve a FIE of ~65,000 which is over 6.5 times larger than a conventional nanorod dimer with the same gap. Meanwhile, the structure maintains a compact size of less than 1/3 the working wavelength. The ASRR configuration also demonstrates superior Purcell factor and fluorescence enhancement. These results demonstrate the ability of ASRR structures for driving breakthroughs in near field responses and can find applications in SERS, fluorescence enhancement, nonlinear optics and enhanced light-matter interactions.

—Supplementary material—

# Subwavelength plasmonic antennas based on asymmetric split-ring-resonators for high near-field enhancements

Yue You, Xiao-Jing Du, Lin Ma, Hua Qiu, Jun He, and Zhong-Jian Yang*

*Hunan Key Laboratory of Nanophotonics and Devices, School of Physics, Central South University, Changsha 410083, China*

*zjyang@csu.edu.cn

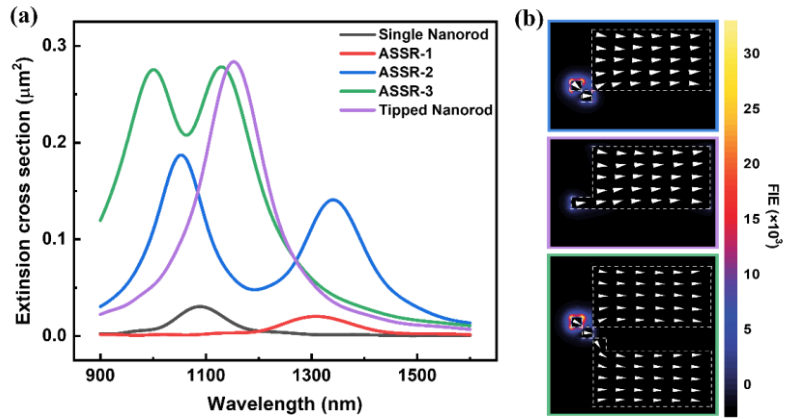

FIG. S1. (a) Extinction spectra of five structures: ASSR-1, ASSR-2, ASSR-3, Nanorod dimer, and ASSR dimer. (b) Linear plot of FIEs and current distribution maps for the ASSR-2, tipped nanorod, and ASSR-3 structures. The color-coded frames correspond to the spectral lines in (a), respectively. The white dashed lines outline the structure contours, and the white arrows indicate the direction of current flow.

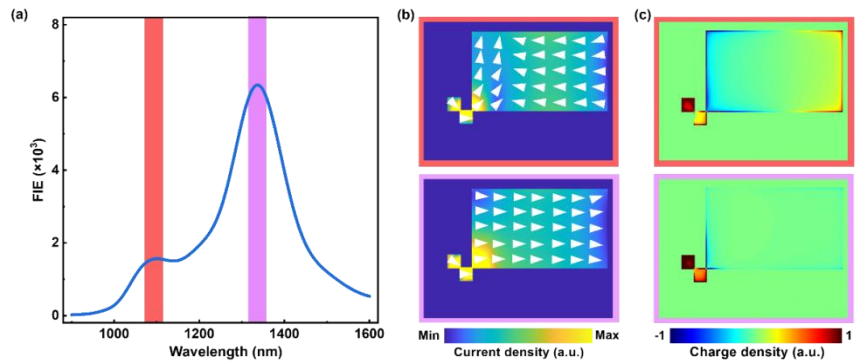

FIG. S2. Higher-order mode of the ASSR-2 structure. (a) FIE spectrum of the ASSR-2 structure at the center of curvature-induced gap point. The red and purple vertical stripes indicate resonance peaks at 1097 nm and 1335 nm, respectively. (b) Current density maps at the 1097 (top) and 1335 nm (bottom). (c) Charge density maps at the 1097 (top) and 1335 nm (bottom).

**Computational methodology and size configuration of matrix antenna**

Numerical simulations were performed using a commercially available finite-difference time-domain (FDTD) solver (Lumerical FDTD). In the simulations, the mesh size surrounding the metallic structures was carefully chosen to $2 \times 2 \times 2 \ nm^3$ to ensure high spatial resolution. The excitation source was implemented as a total-field scattered-field plane wave, and the surrounding medium was assigned a refractive index of $n = 1$ for all simulations. Perfectly matched layers were applied in the $x$, $y$ and $z$ directions to minimize boundary reflections and ensure accurate results. To optimize computational efficiency, the FDTD convergence time was set to 80 fs, a duration sufficient to achieve the desired accuracy for plasmonic systems and enable precise determination of the optical parameters of the structure.

In the design of the antenna, an $s \times s$ matrix was employed, where $s$ is an odd integer, each matrix element $M_{x,y}$ is binary with $M_{x,y} \in \{0,1\}$, and the central element $M_{\frac{s+1}{2},\frac{s+1}{2}}$ is fixed at 0. This matrix serves as a digital representation of a nanostructure, enabling systematic exploration of two-dimensional optical responses under light excitation within a theoretical framework. Balancing the scale of optical response modes with computational time and resource constraints, we selected $s = 21$ to investigate the field intensity enhancement (FIE) of the nanostructure under plane wave excitation.

To map the matrix onto a physical space, every matrix element was modeled as either a $10 \times 10 \times 10 \ nm^3$ gold cube (representing 1) or a same volume of empty area (representing 0), resulting in a $210 \times 210 \times 10 \ nm^3$ gold checkerboard-like complex thin-film nanostructure. Under $x$-polarized plane wave excitation, the FIE spectrum within the central $10 \times 10 \times 10 \ nm^3$ optimization point ($M_{\frac{s+1}{2},\frac{s+1}{2}}$ of the matrix) was systematically studied, providing insights into the localized field enhancement and optical behavior of the designed nanostructure. Specifically, the simulation of an individual structure required approximately 12 minutes.

**Method of creating new individuals**

Various mechanisms within evolutionary algorithms generate individuals from generation $n$ to $n + 1$. These mechanisms form the foundation of evolution optimization (EO) and EoNet [EO and residual convolutional neural network (ResNet)], which differ primarily in the number of individuals produced per generation. For instance, in EO, each generation yields 30 new individuals, whereas EoNet generates approximately 2000 individuals for the subsequent generation.

**Linear crossing:** The encoded genomes of two parental individuals, $A$ and $B$, from generation $n$, produce offspring $O^{(l)}$ in generation $n + 1$. This is achieved by combining the first half of the genome of $A$ with the second half of the genome of $B$, preserving the overall genome length [1,2]. The point $p$ at which the individual genomes are split is chosen randomly. Mathematically, the above can be written as

$$O^{(l)}_{x,y} = \begin{cases} A_{x,y} & x < p \ or \ (x = p \ and \ y \leq p) \\ B_{x,y} & x > p \ or \ (x = p \ and \ y > p) \end{cases}, \qquad (S1)$$

**Spiral crossing:** The encoded genomes of parents $A$ and $B$ from generation $n$, undergo a spiral crossing process to produce offspring $O^{(s)}$ in generation $n+1$. This is achieved by combining the inner part of the genome of $A$ with the outer part of the genome of $B$, preserving the overall genome length. It is important to note that the crossover point $p$ which determines the genomic split, is selected randomly to ensure unbiased genetic variation. Formally, this is expressed mathematically as

$$O^{(s)}_{x,y} = \begin{cases} A_{x,y} & |x - d_c| \leq d \text{ and } |y - d_c| \leq d \\ B_{x,y} & |x - d_c| > d \text{ and } |y - d_c| > d' \end{cases} \quad (S2)$$

where $d_c = \frac{s+1}{2}$, $d = d_c - p$, $s$ represents the scaling factor of the structural matrix.

**Mutation:** Each block of parent $A$ has a 10% probability to toggle between gold and void, resulting in new offspring $O^{(m)}$. This ensures that structures with favorable fitness parameters can maintain their structural patterns, this can be expressed as

$$O^{(m)}_{x,y} = A_{x,y} + (1 - 2 \times A_{x,y}) \times \mathbb{I}_{\{P_{x,y} < 0.1\}}, \quad (S3)$$

where $\mathbb{I}_{\{P_{x,y} < 0.1\}}$ is the indicator function, $P_{x,y}$ denote the probability function defined in the spatial domain. Specifically, when the value of the probability function $P_{x,y}$ at coordinates $(x, y)$ is less than 0.1, the indicator function $\mathbb{I}_{\{P_{x,y} < 0.1\}}$ takes the boolean value 1; otherwise, it takes the value 0.

The probability of applying each method to generate a new individual is uniformly distributed. Additionally, to account for the general structure and the differential impact of the center and edges on optimization parameters in a two-dimensional grid, we implemented a gradient mutation probability ranging from 1% at the center to 5% at the edges for all three generation mechanisms. This approach preserves the fundamental characteristics of the optimized structure while ensuring independence from the parent pool, thereby enabling the evolution process to escape local optima.

**EoNet workflow**

The EoNet workflow involves generating thousands of matrices using evolutionary algorithms. These matrices are directly fed into a neural network to identify those with potential value for electromagnetic simulation calculations. The selected matrices are then converted into structures and computed corresponding FIE values using the FDTD method. Finally, the matrices with the highest fitness values are used as parents to generate offspring for the next generation.

The primary distinction between EO and EoNet lies in the incorporation of a neural network to intervene in the evolutionary process. Therefore, we will provide a detailed description of the network architecture. The neural network employed in EoNet is based on the ResNet-18 architecture, a convolutional neural network with residual connections [3]. By incorporating skip connections that bypass certain convolutional layers, this model effectively mitigates noise in the learning process. The convolutional layers retain the structure of the pre-trained model, while the dense layers are modified to output binary classifications, enabling the determination of whether structures are worth computing for EoNet. Its key advantage over traditional EO lies in its ability to rapidly exclude structures with low-FIE, significantly accelerating the evolutionary process while maintaining the potential for optimization. The efficiency of EoNet's exclusion mechanism directly influences the overall time required for the evolutionary workflow, quantified as the acceleration ratio.

Evaluation metrics play a crucial role in enhancing EoNet's screening efficiency. Based on the confusion matrix for binary classification, the network's results can be categorized into true positives (TP), false positives (FP), true negatives (TN), and false negatives (FN). The network's predictions are denoted as positive ($\omega_1$) or negative ($\omega_0$). TP and TN represent correct judgments, where TP indicates both the actual and predicted outcomes are positive, and TN indicates both are negative. FP and FN represent errors, with FP indicating the false alarm and FN indicating the missed detection. For EoNet, the tolerance for FN errors is significantly lower than for FP errors due to the time-consuming nature of the FDTD simulations. Therefore, the chosen evaluation metric is precision, defined as

$$Precision = \frac{TP}{TP+FP}. \qquad (S4)$$

Decision-making in network predictions directly affects the values within the confusion matrix, thereby influencing precision. As the FIE values increase, EoNet's criteria for judging the computational value of matrices become more stringent, resulting in a naturally imbalanced dataset for binary classification and necessitating minimal decision errors. Bayesian decision theory and cost-sensitive learning can address this challenge. For a sample $X$, precision can be expressed in terms of conditional probabilities

$$Precision = P(X = \omega_1 | \hat{X} = \omega_1) = \frac{P(\hat{X}=\omega_1|X=\omega_1) \cdot P(X=\omega_1)}{P(\hat{X}=\omega_1)} \qquad (S5)$$

where $P(X = \omega_1)$ is the prior probability reflecting data imbalance, $P(\hat{X} = \omega_1 | X = \omega_1)$ is the Recall (sensitivity) and $P(\hat{X} = \omega_1)$ is the marginal probability of predicting the positive class. Precision, as a posterior probability, can be optimized using Bayesian decision rules, the effect shown at Fig. S3. Given the imbalanced nature of the dataset, cost-sensitive learning is employed instead of risk minimization. Considering the cost matrix $C$, where one FDTD calculation takes approximately 12 minutes and the EA generates about 800 matrices per second, the cost matrix is defined as

$$C_{cost} = \begin{bmatrix} C_{11} & C_{12} \\ C_{21} & C_{22} \end{bmatrix} = \begin{bmatrix} 0 & 660s \\ 0.08s & 0 \end{bmatrix}. \qquad (S6)$$

The corresponding risk can be derived from this cost matrix as

$$R(\omega_1|x) = C_{12} \cdot P(y = \omega_0|x) + C_{11} \cdot P(y = \omega_1|x), \qquad (S7)$$
$$R(\omega_0|x) = C_{21} \cdot P(y = \omega_1|x) + C_{22} \cdot P(y = \omega_0|x), \qquad (S8)$$

allowing for the determination of a reasonable decision threshold.

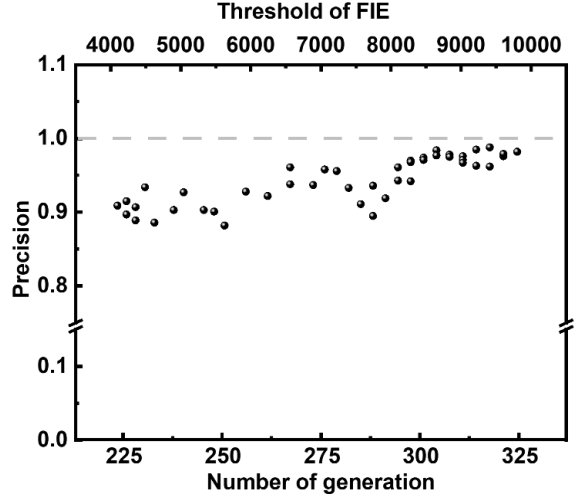

FIG. S3. Evolution of precision across generations in the EoNet.

**Data augmentation**

The learning process of neural networks can be described using Bayesian principles. Let $D$ represent the observed data, $\theta$ the unknown parameters of a model, and assume that all other conditions are encapsulated in $m$. The learning process can be expressed as:

$$P(\theta|D,m) = \frac{P(D|\theta,m) \cdot P(\theta|m)}{P(D|m)} \tag{S9}$$

where $P(D|\theta,m)$ is the likelihood of parameters $\theta$ in model $m$, $P(\theta|m)$ is the prior probability of $\theta$, and $P(\theta|D,m)$ is the posterior probability of $\theta$ given data $D$. Learning involves transforming prior knowledge or assumptions about the parameters $P(\theta|m)$ into posterior knowledge $P(\theta|D,m)$ through data $D$ [4]. The core of this learning process lies in $P(D|\theta,m)$, where designing the likelihood is crucial for enhancing learning capability.

One effective method to improve learning performance is data augmentation, which involves increasing the quantity of data $D$ to extend the learning process. Data augmentation aims to enhance the generalization capability of the data while preserving its information entropy. For two-dimensional plasmonic structures, the core information includes the structural configuration and the central optimization point. Conventional data processing techniques can easily disrupt this structural information entropy, introducing noise and reducing learning effectiveness. Therefore, it is essential to first apply feature-preserving processing to the images to retain critical data characteristics. For instance, by marking the central optimization point position in dataset, the structural optimization point and configuration can be preserved throughout data processing, maintaining information entropy. As illustrated in Fig. S4, we choose to highlight the central optimization point position in red and extend it slightly along the plane of the electric field of the incident light to indicate the polarization direction, forming a near-rhomboid structure. Subsequently, data augmentation techniques are applied to these feature-marked data, including center cropping, random cropping, horizontal flipping, vertical flipping, random rotation, color jittering, Gaussian blurring, random perspective, affine transformations, and elastic deformations. These techniques expand the dataset by a factor of eleven, significantly enhancing the model's generalization capability. It is crucial to ensure that the structural features remain preserved throughout these augmentations.

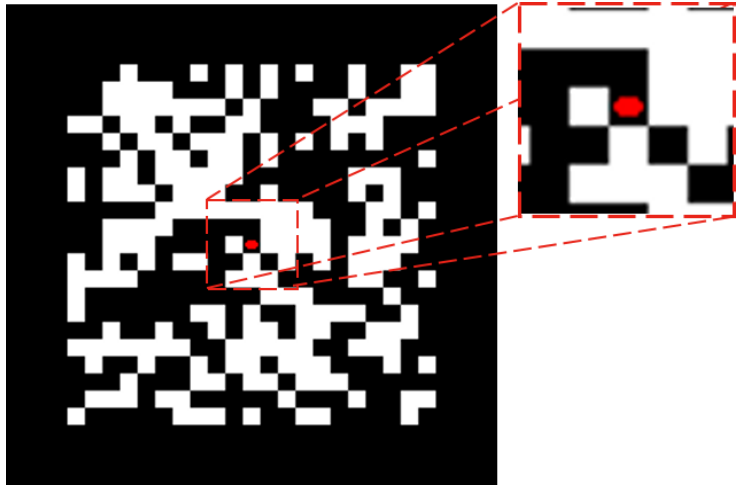

FIG. S4. Data processing schematic. The structure is represented as a black-and-white matrix image, where black corresponds to empty regions and white represents the physical structure. A zoomed view highlights the central region of the structure, with a near-rhomboid red marking in the optimization point for enhanced visibility.

**Occam's Razor algorithm**

The Occam's Razor algorithm is designed to optimize complex structures and identify intrinsic response patterns through structural simplification [5]. It employs an iterative process to eliminate components that do not contribute to the target parameters, retaining only the core structures that significantly influence these parameters. Considering the connectivity of current flow and computational resource consumption, optimization is achieved by trimming edge structures. Specifically, edge blocks are identified where the nearest neighbors in the $x$ and $y$ directions are not fully occupied by Au cubes, i.e., $M_{i\pm1,j} \cap M_{i,j\pm1} = 0$.

In each step, the impact of virtually removing each edge block on the central FIE value is recorded (Fig. S5). This process, referred to as an edge-removing loop, is repeated for all edge blocks from $M_{1,1}$ to $M_{21,21}$. Subsequently, a greedy strategy is employed to permanently remove up to 10% of the blocks that, when virtually removed, most positively impact the FIE value. This removal of non-contributing blocks simplifies the structure. The entire process is repeated until the FIE value of the original structure is reduced to 90% of its initial value, at which point the loop terminates.

The optimization of a two-dimensional plasmonic matrix antenna using the Occam's Razor algorithm is illustrated in Fig. S5(a) and the stable FIE value across iterations demonstrates the algorithm's effectiveness in optimizing complex structures. In Fig. S5(b), the edge block removal process is analyzed for selected iterations. Benchmarking FIE values against the pre-loop structure reveals that central structures significantly impact FIE, highlighting their critical role. As iterations progress, blocks with positive FIE changes decrease in number and impact, turning entirely negative by the fourteenth iteration. Connector blocks between sections also significantly affect FIE, with more influential connectors observed in the fourteenth iteration compared to the first, indicating convergence toward the optimal response pattern.

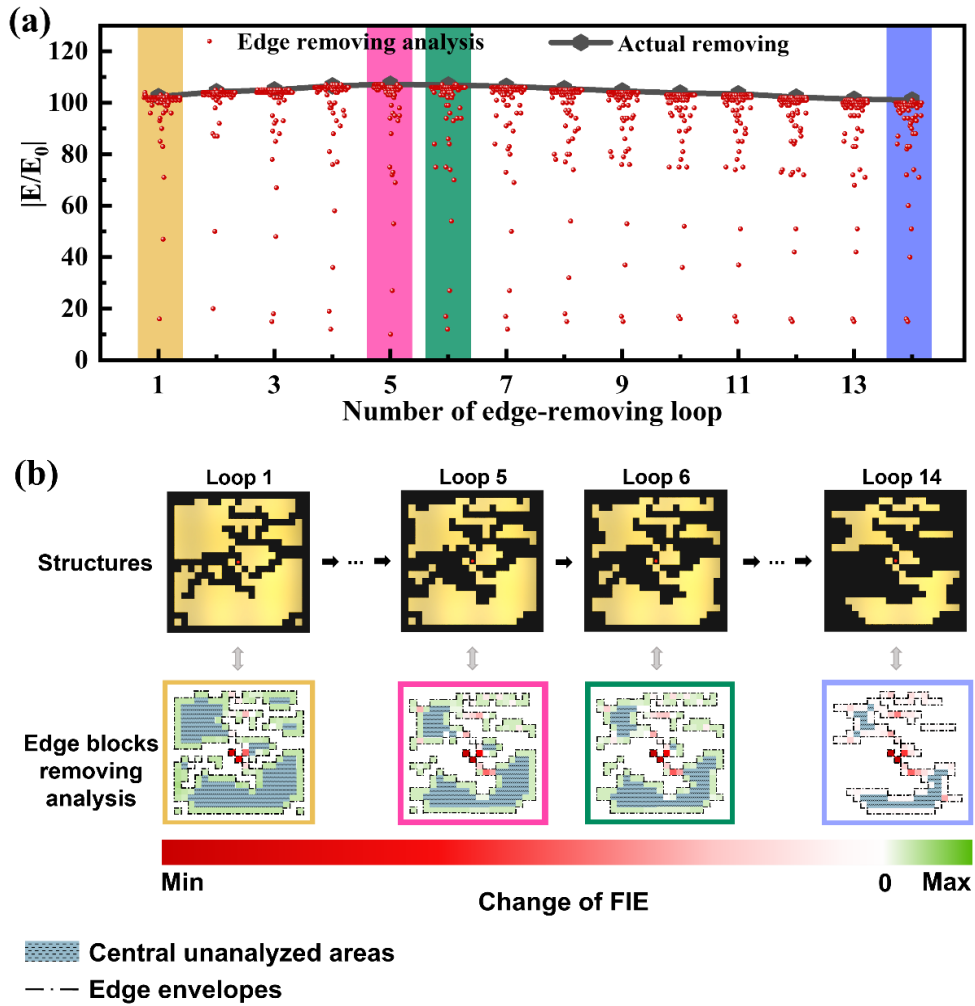

FIG. S5. The Occam's Razor optimization process. (a) The evolution of the Occam's Razor algorithm's structural edge looping process. The black symbol-line represents the FIE value evolution for the actual removed structures, while the red dots indicate the impact of virtually removing blocks on the FIE value. For clarity, the red dots within each loop are spaced uniformly along the horizontal axis to maintain consistent intervals between adjacent loops. (b) The upper panel displays structural diagrams for Loop 1, Loop 5, Loop 6, and Loop 14. The lower panel shows the corresponding changes in FIE value at the optimization position when each single edge block is removed.

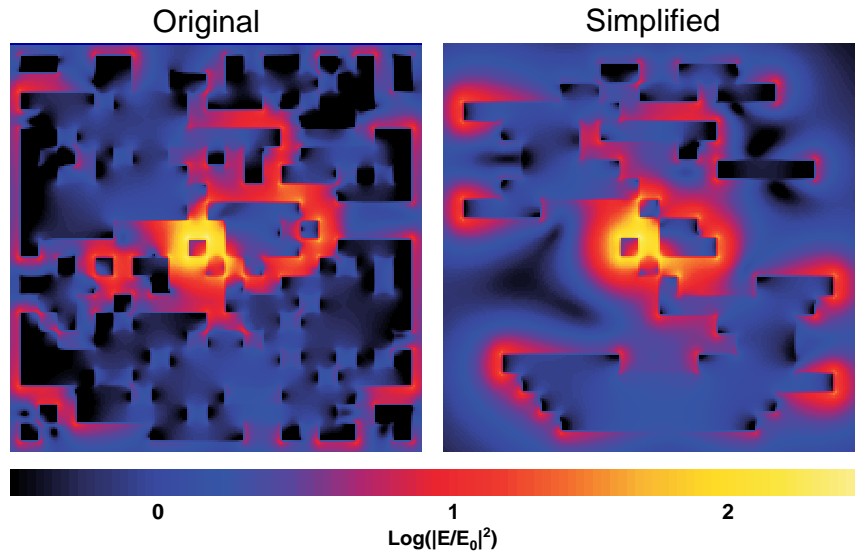

FIG. S6. Logarithmic FIE distributions of original and simplified structures. The structures are the same as that in Fig. 3 in the main text.

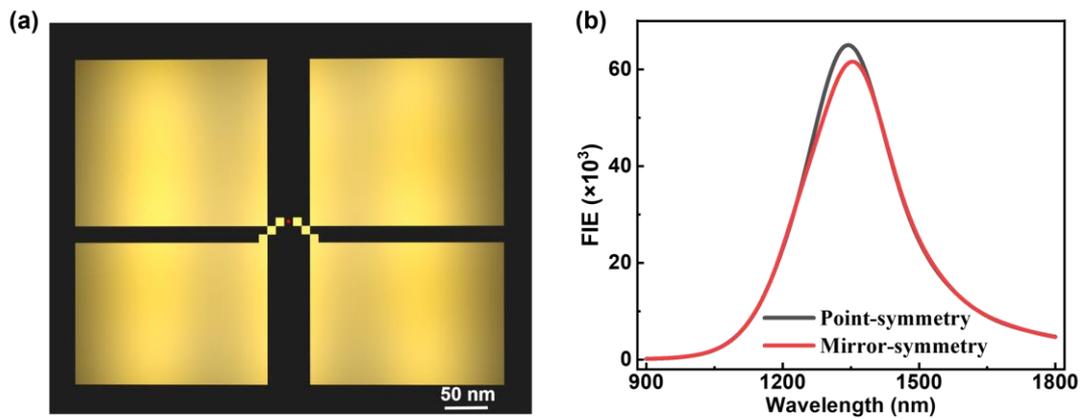

FIG. S7. (a) A ASRR dimer with mirror-symmetric configuration. (b) FIE spectra of point-symmetric and mirror-symmetric configuration for ASRR dimer. The point-symmetric dimer is the same as that in Fig. 4 in the main text.

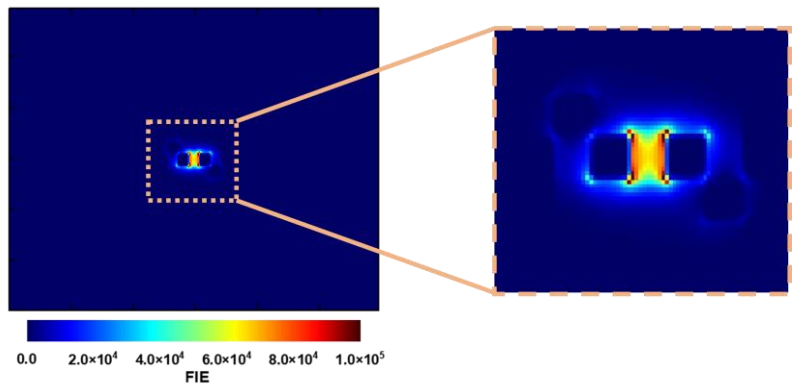

FIG. S8. Linear FIE distribution of an ASRR dimer and the zoomed view of center region. The dimer is the same as that in Fig. 4 in the main text.

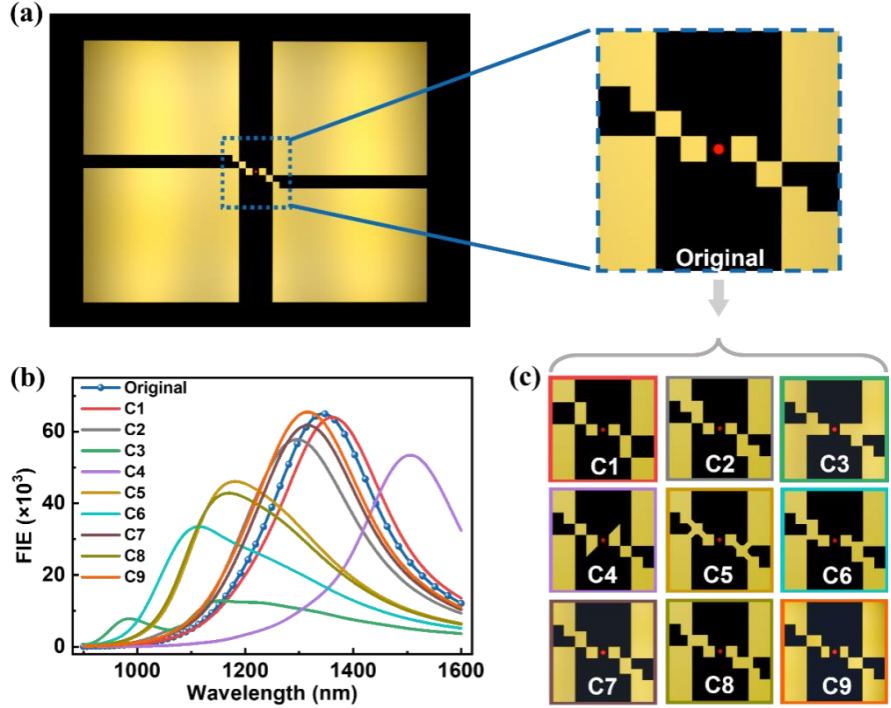

FIG. S9. Influence of the central geometries on FIE values. (a) The ASSR dimer with zoomed view of the central $90 \times 90\,nm^2$ region. The red dot indicates the location of the optimization point for monitoring the FIE value. (b) The FIE spectra of structures with different geometries around the center region. The geometries (from C1 to C9) are shown in (c).

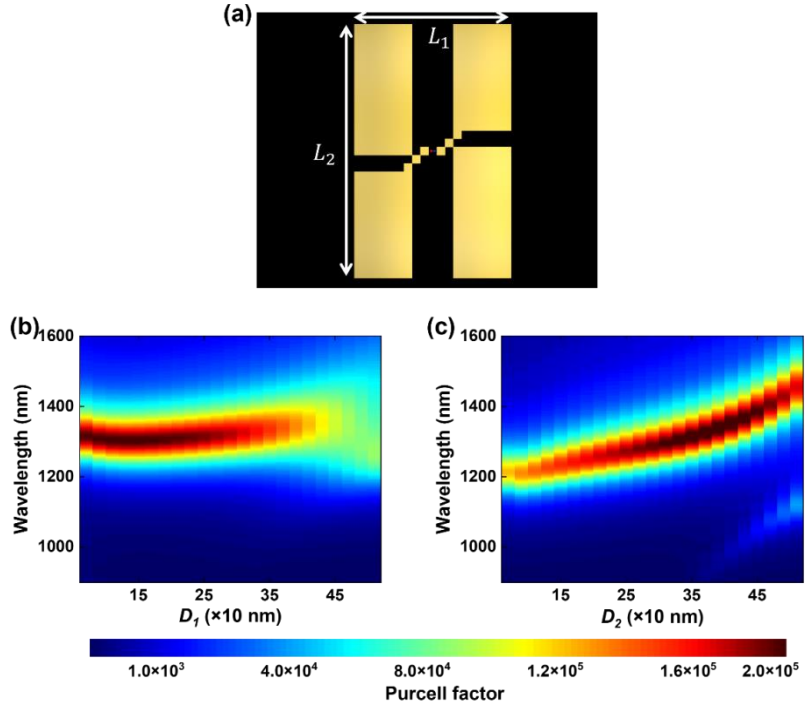

FIG. S10. (a) Schematic of an ASSR dimer structure, with a dipole source positioned at the gap center. Purcell factors of ASSR dimers with different sizes $L_1$ and $L_2$ are shown in (b) and (c), respectively. The gap center region of the dimer is the same as that in Fig. 5 in the main text.

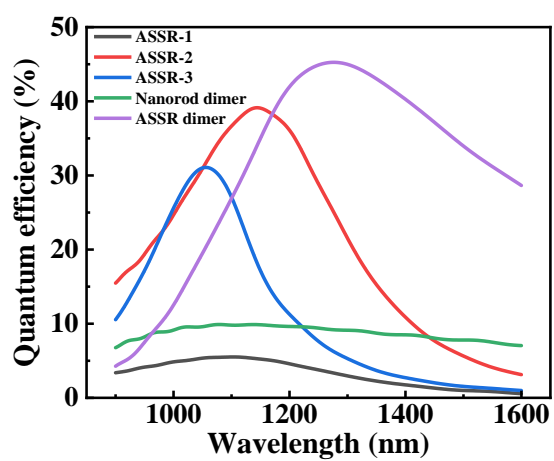

FIG. S11. Quantum efficiency spectra of different structures. The structures ASSR-1, ASSR-2, ASSR-3, nanorod dimer, and ASSR dimer correspond to those shown in Fig. 2 and 4 in the main text.